\documentclass[aps,prl,twocolumn,showpacs,superscriptaddress,floatfix]{revtex4-1}
\usepackage{amsmath,amssymb}
\usepackage{graphicx}
\usepackage{epsfig}
\usepackage{color}
\usepackage{rotating}
\usepackage{bm}
\usepackage{setspace}
\usepackage{hyperref}

\begin{document}
\title{Probing Dirac Fermion Dynamics in Topological Insulator Bi$_2$Se$_3$ Films \\ with Scanning Tunneling Microscope}
\author{Can-Li Song}
\email[]{clsong07@mail.tsinghua.edu.cn}
\affiliation{State Key Laboratory of Low-Dimensional Quantum Physics, Department of Physics, Tsinghua University, Beijing 100084, China}
\affiliation{Collaborative Innovation Center of Quantum Matter, Beijing 100084, China}
\author{Lili Wang}
\author{Ke He}
\author{Shuai-Hua Ji}
\author{Xi Chen}
\author{Xu-Cun Ma}
\email[]{xucunma@mail.tsinghua.edu.cn}
\author{Qi-Kun Xue}
\affiliation{State Key Laboratory of Low-Dimensional Quantum Physics, Department of Physics, Tsinghua University, Beijing 100084, China}
\affiliation{Collaborative Innovation Center of Quantum Matter, Beijing 100084, China}
\date{\today}

\begin{abstract}
Scanning tunneling microscopy and spectroscopy have been used to investigate the femtosecond dynamics of Dirac fermions in the topological insulator Bi$_2$Se$_3$ ultrathin films. At two-dimensional limit, bulk electrons becomes quantized and the quantization can be controlled by film thickness at single quintuple layer level. By studying the spatial decay of standing waves (quasiparticle interference patterns) off steps, we measure directly the energy and film thickness dependence of phase relaxation length $l_{\phi}$ and inelastic scattering lifetime $\tau$ of topological surface-state electrons. We find that $\tau$ exhibits a remarkable $(E-E_F)^{-2}$ energy dependence and increases with film thickness. We show that the features revealed are typical for electron-electron scattering between surface and bulk states.
\end{abstract}
\maketitle
\begin{spacing}{1.02}
Topological insulators (TIs) are generally characterized by a bulk band gap and an odd number of topologically protected metallic surface states \cite{moore2010birth, hasan2010colloquium, qi2011topological}. Strong spin-orbit interaction brings about many exotic physical properties in TIs and ensures the robustness of the surface metallic states against non-magnetic impurities. With the nontrivial spin textures of surface states, three-dimensional (3D) TIs in reduced dimensions have generated tremendous interests since their discovery, due to the enhanced surface-to-volume ratio and potential for developing future dissipationless spintronics \cite{kong2009topological, zhang2010crossover, Liu2011electron, kim2011thickness, Taskin2012manifestation, bansal2012thickness, chang2013experimental}. Angle-resolved photoemission spectroscopy (ARPES) on Bi$_2$Se$_3$ films has revealed an energy gap opening associated with the hybridization of top and bottom surface states in the two dimensional (2D) limit \cite{zhang2010crossover}. Extensive transport experiments have demonstrated thickness dependent electrical resistivity and magnetoresistance in ultrathin films of TIs , which has been suggested as the combined effects of strong electron-electron ($e$-$e$) interactions and topological delocalization \cite{Liu2011electron, kim2011thickness, Taskin2012manifestation, bansal2012thickness}. On the other hand, ARPES can act as a powerful tool to probe directly the dynamics of Dirac fermions \cite{Higashiguchi2007}. So far as we know, only a few such experiments were reported, but focused primarily on electron-phonon ($e$-$p$) and electron-impurity scattering and led to inconsistent results \cite{Hatch2011, Pan2012, Chen2013}. Thus far, the $e$-$e$ interactions of topological surface states in TIs have scarcely been addressed separately and established conclusively.

The phase relaxation length $l_{\phi}$, defined as the average distance a quasiparticle can propagate without losing its phase memory, is usually thought of as a key quantity to describe the femtosecond dynamics of electrons in solid state physics including TIs. Therefore an accurate estimate of $l_{\phi}$ in TIs is highly desirable, but challenging in experiment. In Bi$_2$Se$_3$ films $l_{\phi}$ at the Fermi level ($E_F$) was recently studied by magnetoresistance measurements \cite{Liu2011electron, kim2011thickness, Taskin2012manifestation, bansal2012thickness}, but the obtained results of the intrinsic $l_{\phi}(E_F)$ as well as its thickness dependence are not consistent. This may stem from varying surface defect scattering (i.e. step edges and residual impurities) in the samples from different groups. An alternative access to inelastic lifetime $\tau$, and hence to $l_{\phi}$ = $v_F\tau$ ($v_F$ is the group velocity of electrons), has recently become available by analyzing quantitatively the energy widths of the quantized topological surface states with a scanning tunneling microscope (STM) \cite{seo2010transmission}. Such local technique eliminates the residual scattering and allows the estimate of $l_{\phi}$ at various energies. However, the method demands elaborate modeling to remove the instrumental effects, which also severely limits an accurate determination of $l_{\phi}$.

In this Letter, we study the decay of standing waves (SWs) off descending straight step edges in Bi$_2$Se$_3$ films [Fig.\ 1(a)] with scanning tunneling microscopy and spectroscopy (STS). We measure directly $l_{\phi}$ (or equivalently $\tau$) as a function of energy and film thickness. All experiments are carried out on a Unisoku ultra-high vacuum STM system equipped with molecular beam epitaxy (MBE) for film growth. High-quality Bi$_2$Se$_3$ films with controlled thickness are grown by co-evaporating high-purity Bi (99.999$\%$) and Se (99.999$\%$) onto graphitized 6$H$-SiC(0001) substrate kept at $220^{\circ}$C \cite{zhang2010crossover, song2010topological}, except for 19 quintuple layer (QL) film at $260^{\circ}$C. Prior to the STM/STS measurements at 4.8 K, a polycrystalline W tip is cleaned by electron-beam heating in MBE chamber, and then calibrated with STS of Pb films grown on a Si(111) substrate in an energy range of -1.0 eV $\sim$ 1.5 eV \cite{Ma2007}. All the differential conductance $dI/dV$ spectra and maps are acquired using a standard lock-in technique with a small bias modulation of 10 mV at 987.5 Hz.

\end{spacing}
\begin{figure}[tb]
\includegraphics[width=\columnwidth]{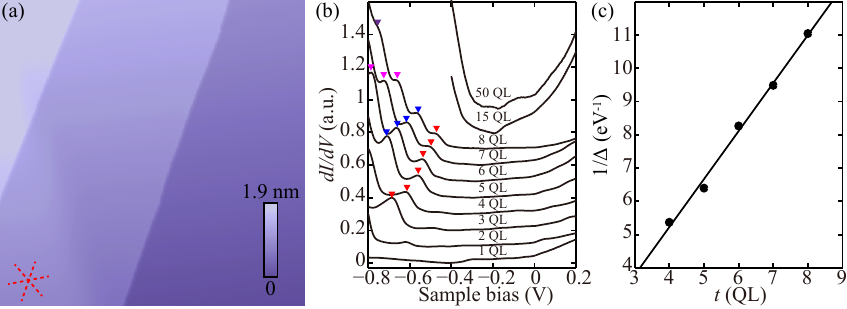}
\caption{(a) STM topographic image of Bi$_2$Se$_3$ film with a nominal thickness of 15 QL ($V_{\mathrm{s}}$ = 4.0 V, $I$ = 0.1 nA, 200 nm $\times$ 200 nm). The step edges are invariably oriented along the closed-packed directions of Bi$_2$Se$_3$, indicated by the three red dashes. (b) A series of $dI/dV$ curves showing the quantum quantization of electrons (triangles) in the 2D limit. The spectra have been vertically offset for clarity. Set point: $V_{\mathrm{s}}$ = 0.2 V, $I$ = 0.1 nA. (c) Inverse of the energy separation between the quantized states marked by red and blue triangles in (b) as a function of film thickness $t$.
}
\end{figure}

Figure 1(b) depicts a series of differential conductance $dI/dV$ spectra in the as-grown Bi$_2$Se$_3$ films with various thicknesses ranging from 1 QL up to 50 QL. Above $\sim$10 QL, the spectra display a minimum at $\sim$0.2 eV, which corresponds to the Dirac point of topological surface states and agrees with those reported previously \cite{Cheng2011Landau, Hanaguri2010Monmentum}. As the films reach the 2D limit, however, additional discrete states (marked by colored triangles) are discernible in the valence band and alter in energy position with film thickness $t$. These are reminiscent of the M-shaped 2D electron gas observed by ARPES \cite{zhang2010crossover, bianchi2010coexistence}, which has been contentiously interpreted as quantum well states (QWSs) \cite{zhang2010crossover, bianchi2010coexistence, bahramy2012emergent} or the expansion of van der Waals' spacing caused by impurity intercalation \cite{eremeev2012effect}. Our experiments are conducted on MBE-grown clean films and little suffers from impurity intercalation, and the results supports strongly the QWSs scenario. This is further evidenced by revealing the film thickness $t$ dependent energy separation ($\Delta$) between adjacent QWSs, plotted in Fig.\ 1(c): the linear relationship between 1/$\Delta$ and $t$ demonstrates indubitably that all the discrete states observed here and also by previous ARPES experiments originate merely from the quantization of bulk electron states \cite{altfeder1997electron}. Considering the vanishing QWSs in thick Bi$_2$Se$_3$ films [Fig.\ 1(b)], we note that the quantization of electron in Bi$_2$Se$_3$  films and bulk single crystals are driven distinctively and primarily by the dimensionality and band-bending effects \cite{zhang2010crossover, bianchi2010coexistence, bahramy2012emergent}, respectively.

\begin{figure*}[tb]
\includegraphics[width=2\columnwidth]{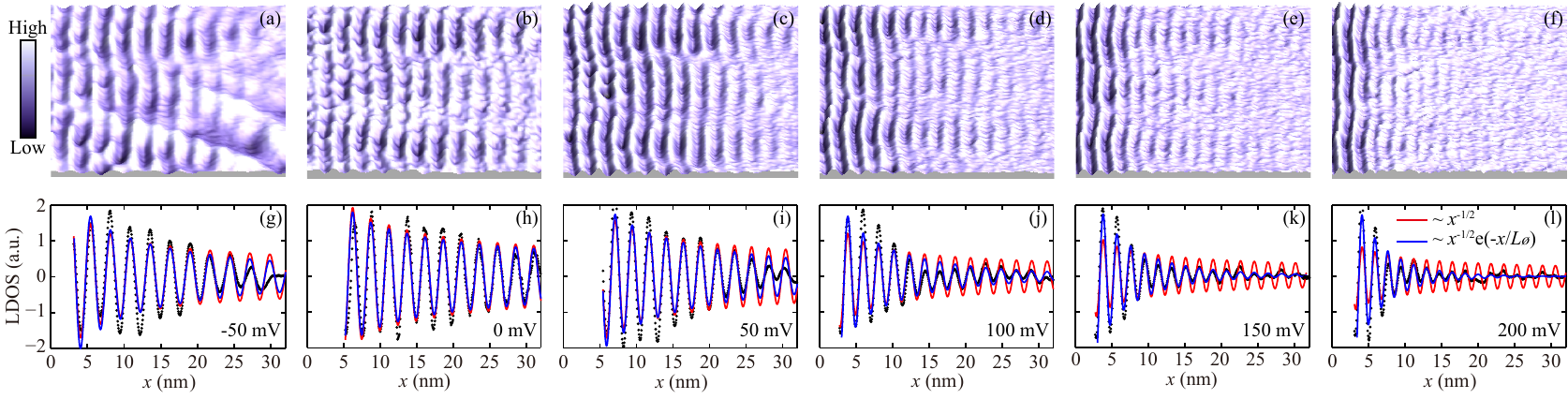}
\caption{(a-f) LDOS maps of 1 QL Bi$_2$Se$_3$ film taken on the upper terrace near a step edge ($I$ = 0.1 nA, 30 nm $\times$ 30 nm), and (g-l) the averaged LDOS (black dots) as a function of distance $x$ from the step edge for various energies. Note that the mean value of each LDOS map has been subtracted. The SWs patterns with different decay rate are clearly evident. Blue and red lines show the best fits of the decaying LDOS with and without a loss of phase coherence considered, respectively.
}
\end{figure*}

\begin{figure*}[tb]
\includegraphics[width=2\columnwidth]{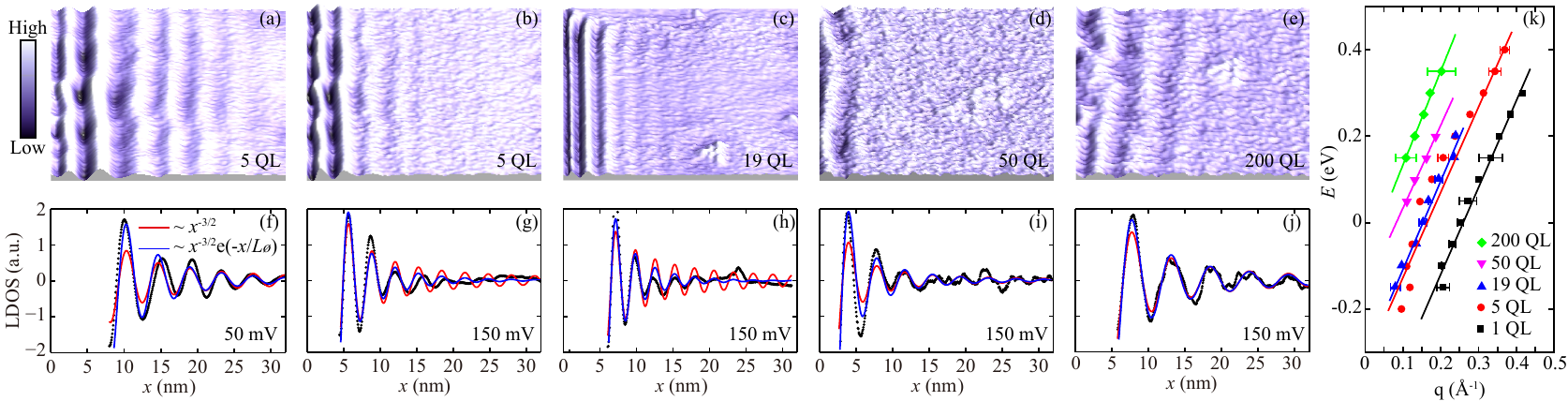}
\caption{(a-e) LDOS maps ($I$ = 0.1 nA, 30 nm $\times$ 30 nm), and (f-j) the averaged LDOS (black dots) as a function of distance $x$ from the step edge for various film thicknesses and energies. Blue and red lines show the best fits of LDOS with and without a loss of phase coherence considered, respectively. (k) Energy $E$-q dispersion relation deduced by fitting the averaged LDOS around the step edges, with q indicating the scattering wave vector. The Dirac points for Bi$_2$Se$_3$ films from 1 QL to 200 QL  are estimated to be -0.51 eV, -0.31 eV, -0.30 eV, -0.16 eV and -0.08 eV by intercept of the $E$-q dispersions with the energy axis at q = 0.
}
\end{figure*}

In order to measure the phase coherence length $l_{\phi}$, we select straight step edges in an almost defect-free region ($\sim$100 nm $\times$ 100 nm) to minimize unwanted electron scattering. On the upper terrace, we acquire simultaneously the topographic image (shown in Fig.\ S1)\cite{supplementary} and differential conductance $dI/dV$ map in an area of 30 nm $\times$ 30 nm, similar to the previous reports \cite{Zhang2009Experimental, Wang2011Power}. Here care should be taken to determine the sample local density of states (LDOS) from which $l_{\phi}$ is extracted. In the vicinity of step edges, strong spatial oscillation of SWs alters the tip-sample distance, and consequently the measured $dI/dV$ will not exactly represent the LDOS. To estimate $l_{\phi}$ more accurately, we have recovered the LDOS from STM topography and $dI/dV$ map using a well-tested algorithm \cite{li1997local}, as illustrated in Figs.\ 2(a-f) and 3(a-e). Here the work functions of W tip $\phi_t$ = 4.55 eV and Bi$_2$Se$_3$ $\phi_s$ = 5.5 eV \cite{wu2013tuning} are adopted. Further variation of $\phi_t$ + $\phi_s$ up to 0.5 eV is found to little affect the reconstructed LDOS. Moreover, a significantly small lock-in bias modulation of 10 meV was adopted to reduce its additional decay in \textit{dI/dV} map, which may complicate the estimate of $l_{\phi}$ \cite{burgi1999probing}. By slightly rotating the LDOS maps to align the step edge vertically, we average the LDOS as a function of distance $x$ from the step edge, as illustrated in Figs.\ 2(g-l) and 3(f-j).

\begin{figure}[b]
\includegraphics[width=\columnwidth]{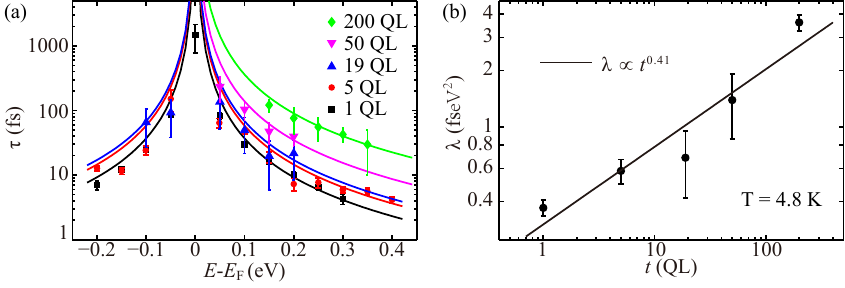}
\caption{(a) Inelastic lifetime $\tau$ of topological Dirac fermions in Bi$_2$Se$_3$ as a function of energy and film thickness. Colored lines depict the best fits of lifetime to $\tau$ = $\lambda (E-E_F)^{-2}$ for various film thicknesses. (b) Double-logarithmic plot of the extracted $\lambda$ from (a) versus film thickness $t$. $\lambda$ increases as $t^{0.41\pm0.08}$ with film thickness $t$.
}
\end{figure}

As expected, the SWs are clearly visible on the terrace and decay as a function of $x$. In ordinary 2D electron gas \cite{burgi1999probing, jeandupeux1999thermal, Crampin2005phase}, it has already been demonstrated that the decay is primarily governed by the combined effects of geometric factor ($\sim$$1/x^{0.5}$) and phase decoherence ($\sim$e$^{-x/l_{\phi}}$), whereas the latter has completely been neglected in TIs by assuming $l_\phi=\infty$ \cite{Wang2011Power, alpichshev2010stm}. Our attempts to fit the LDOS by the predicted cos(q$x+\phi$)$/x^{\alpha}$ fail if the loss of phase coherence is not considered (red curves in Figs.\ 2 and 3), in particular for those at energies far away from $E_F$. The observed LDOS invariably exhibit a faster decay than $1/x^\alpha$, where $\alpha$ = 0.5 and 1.5 are found to best represent the LDOS for $t$ = 1 and for $t$ $\geqslant$ 5, respectively (here q and $\phi$ denote the scattering wave vector and phase shift, respectively). Our careful examination shows better fits of the LDOS if the phase decoherence effect is included (blue curves in Figs.\ 2 and 3), namely
\begin{equation}
\textrm{LDOS}(\textrm{q}, x) \varpropto |r|\frac{\textrm{cos}(\textrm{q}x+\phi)}{x^{\alpha}}e^{-x/l_\phi},
\end{equation}
where $r$ indicates the electron reflection coefficient by the step-edge barrier \cite{Crampin2005phase}. For a finite barrier (e.g. step edge), $|r|$ is generally smaller than 1, but it does not affect the estimate of q and $l_{\phi}$. Our observation thus presents a compelling evidence that the phase coherence has been appreciably lost in the electron scattering process.

Now we comment on the different geometric parameter $\alpha$ between 1 QL and thicker ($t$ $\geqslant$ 5) Bi$_2$Se$_3$ films. In TIs with weak warping effect, such as Bi$_2$Se$_3$ investigated here, the unusual topological spin texture prohibits backscattering, leading to a faster decay of SWs off steps as $1/x^{1.5}$ \cite{Wang2011Power, liu2012stationary}, in stark contrast to the $1/x^{0.5}$ decay in an ordinary 2D electron gas \cite{burgi1999probing, jeandupeux1999thermal}. This matches excellently with our observations for thicker Bi$_2$Se$_3$ films in Figs.\ 3 (a-j). For single QL Bi$_2$Se$_3$ film, however, the top and bottom surface states can talk with each other and break the momentum-spin helical locking symmetry of topological surface states, confirmed recently by spin-polarized ARPES experiments \cite{Landolt2014Spin, neupane2014observation}. As a consequence, the backscattering from the top surface state is allowed, leading to the $1/x^{0.5}$ decay in single QL Bi$_2$Se$_3$ film [Fig.\ 2], similar to the ordinary 2D electron gas. Therefore, our simple scattering experiments off steps have unequivocally demonstrated not only that the backscattering of topological surface states by nonmagnetic impurities (e.g. step edges) is indeed suppressed \cite{Zhang2009Experimental, roushan2009topological}, but also that the top and bottom surface states hybridize strongly and break the chiral spin texture in the 2D limit \cite{zhang2010crossover, Landolt2014Spin, neupane2014observation}. More significantly, the phase decoherence effect on the SWs decay is well separated from the geometric factor in Eq.\ (1), allowing a more accurate absolute estimate of $l_{\phi}$. Figure 3(k) summarizes the $E$-q dispersion relation of Bi$_2$Se$_3$ surface states, where the wave vector q are extracted from the fits of LDOS to Eq.\ (1). Colored lines reveal the linear fits to the data for various film thicknesses. The approximately equal slope hints at the same dominant scattering channel for various film thicknesses. Using q = 2k \cite{Wang2011Power}, we deduce the Fermi velocity to be $v_F$ = $4.0\pm0.2$ eV$\cdot${\AA} along the $\Gamma$-\textit{M} direction.

Figure 4(a) plots the inelastic lifetime $\tau$, calculated via  $\tau = l_{\phi}/v_F$, as a function of energy and film thickness. The error bars primarily indicate the standard derivation of $\tau$ values obtained from the LDOS fits over various $x$ ranges. Note that in our measurements we have used a typically small tunneling current of $I$ = 0.1 nA, at which one excited electron about every 1.6 ns ($\gg\tau$) is injected into ($V_{\mathrm{s}}>0$) or knocked out of ($V_{\mathrm{s}}<0$) the samples. This avoids nonequilibrium probe and electron cascade effect. It is clearly evident from Fig.\ 4(a) that the inelastic lifetimes $\tau$ of topological Dirac fermions are strongly reduced when the energy $E$ departs away from $E_{F}$. This is typical for $e$-$e$ scattering, since the $e$-$p$ lifetimes are generally little dependent of quasiparticle energy for the small energy range studied here. Moreover, we find that $\tau$ changes with energy in a power-law manner. In Fig.\ 4(a) the five colored lines shows the best fits of $\tau$ for various Bi$_2$Se$_3$ films by
\begin{equation}
\tau =  \frac{\lambda}{(E-E_F)^{2}}.
\end{equation}
Such power law behavior agrees well with 3D interacting Fermi liquid theory for electron-hole pair creation \cite{nozieres1999theory}, similar to previous observation for conventional surface states on Ag(111) and Cu(111) \cite{burgi1999probing}. We thus believe that the phase relaxation lengths $l_{\phi}$ or lifetimes $\tau$ measured here are predominantly governed by inelastic $e$-$e$ scattering in Bi$_2$Se$_3$ films. This is understandable since $e$-$p$ scattering is expected to be insignificant at the low temperature of 4.8 K.

Finally we shed more insight into the observed $e$-$e$ scattering by studying the film thickness $t$ dependence of $\lambda$, plotted in Fig.\ 4(b). Here a larger $\lambda$ means a longer phase coherence length $l_{\phi}$ or lifetime $\tau$. It is found that $\lambda$ increases with film thickness as $\lambda$ $\varpropto$ $t^{0.41\pm0.08}$. If the inelastic scattering proceeds mostly via the 2D channel, i.e. the intraband scattering within the surface states, $\lambda$ would depend little on $t$ since the surface should not alter with film thickness. Therefore, the film thickness-dependent inelastic lifetime $\tau$ (or $\lambda$) reveals directly a predominant 3D scattering channel in Bi$_2$Se$_3$ TIs, namely $e$-$e$ scattering between surface and bulk states. This is consistent with the power law dependence of $\tau$ on energy observed above, predicted in 3D interacting Fermi liquid theory. The enhanced $e$-$e$ scattering in thinner film is primarily related with reduced dimensionality, which results into weaker electron screening and hence stronger Coulomb interaction between electrons. As compared to the $e$-$e$ scattering in Ag and Cu \cite{burgi1999probing}, $\lambda$ appears substantially smaller in Bi$_2$Se$_3$, indicative of weaker electron screening in TIs than that in ordinary metals.

In summary, our detailed STM and STS study of the SWs decay in MBE-grown Bi$_2$Se$_3$ films has revealed the significant loss of phase coherence. By fitting the LDOS, we have estimated the energy and film thickness dependence of the phase coherence length $\l_{\phi}$ or inelastic lifetime $\tau$. The inelastic lifetime $\tau$ depends on both energy and film thickness in a power law way. This has convincingly demonstrated the significance of $e$-$e$ scattering between surface and bulk states in TIs. The present STM/STS approach avoids macroscopic integral measurements, allowing for a better estimate of $\l_{\phi}$ and $\tau$. It should be highly desirable to generalize this technique to other TIs, and also to probe the temperature dependence of SWs from which the contributions of $e$-$p$ scattering on the phase decoherence can be quantified.

\begin{acknowledgments}
This work was supported by National Science Foundation and Ministry of Science and Technology of China. All STM image and LDOS maps were processed by Nanotec WSxM software \cite{horcas2007wsxm}.
\end{acknowledgments}
%

\end{document}